\documentclass[twocolumn,prc,superscriptaddress,showpacs]{revtex4}
\usepackage{float}
\usepackage{amsmath}
\usepackage{graphicx}
\usepackage{amssymb}

\newcommand{\simge}{\hspace*{0.2em}\raisebox{0.5ex}{$>$}
     \hspace{-0.8em}\raisebox{-0.3em}{$\sim$}\hspace*{0.2em}}
\newcommand{\simle}{\hspace*{0.2em}\raisebox{0.5ex}{$<$}
     \hspace{-0.8em}\raisebox{-0.3em}{$\sim$}\hspace*{0.2em}}

\makeatletter

\providecommand{\tabularnewline}{\\}

\makeatother
\begin{document}

\title{Effective Theory for Trapped Few-Fermion Systems}

\author{I. Stetcu}

\affiliation{Theoretical Division, Los Alamos National Laboratory, Los Alamos,
New Mexico 87545, USA}

\author{B.R. Barrett }

\affiliation{Department of Physics, University of Arizona, Tucson, Arizona 
85721, USA}

\author{U. van Kolck}

\affiliation{Department of Physics, University of Arizona, Tucson, Arizona 
85721, USA}

\affiliation{Kernfysisch Versneller Instituut, Rijksuniversiteit Groningen, 
Zernikelaan 25, 9747 AA Groningen, The Netherlands}

\affiliation{Instituto de F\'{\i}sica Te\'{o}rica, 
Universidade Estadual Paulista,
Rua Pamplona 145, 01405-900 S\~{a}o Paulo, SP, Brazil}

\author{J.P. Vary}

\affiliation{Department of Physics and Astronomy, Iowa State University, Ames,
Iowa 50011, USA}

\date{May 29, 2007}

\preprint{LA-UR-07-3398}

\begin{abstract}
We apply the general principles of effective field theories to the
construction of effective interactions suitable for few- and many-body
calculations in a no-core shell model framework. 
We calculate the spectrum of systems with
three and four 
two-component fermions in a harmonic trap.
In the unitary limit, we find that three-particle results are 
within 10\% of known semi-analytical values even in small model spaces. 
The method is very general, and can be readily extended to other regimes,
more particles, different species (e.g., protons and neutrons in nuclear
physics), or 
more-component fermions (as well as bosons). 
As an illustration, we present calculations 
of the lowest-energy three-fermion states away from the unitary limit 
and find a possible inversion of parity in the ground state 
in the limit of trap size large compared to the scattering length.
Furthermore, we investigate the lowest positive-parity states for
four fermions,
although we are
limited by the dimensions we can currently handle in this case. 
\end{abstract}
\pacs{03.75.Ss, 34.20.Cf, 21.60.Cs}

\maketitle

\section{Introduction}

The properties of strongly interacting Fermi gases have been the object
of great interest in recent years. 
Feshbach resonances allow the tuning of the interaction between 
trapped particles so that one can study the evolution from 
a dilute Fermi gas to a Bose-Einstein condensate. 
In three-dimensional optical lattices one can reach the
low-tunneling regime
where each site is an essentially isolated harmonic trap 
occupied by few fermions \cite{atomexpt,stoferle:030401}.  
This opens up a new window into the study of few-body systems with
two-body scattering lengths $a_2$ that are large compared to 
the range $r_0$ of the interaction, $|a_2|\gg r_0$.

Large-$|a_2|$ systems are of particular theoretical interest
when particle momenta $Q$ are small compared to $1/r_0$,
because then they exhibit
universal behavior, that is, the system properties depend essentially
only on $a_2$ (and for bosons or multi-component fermions, also
on a three-body parameter), but not on the details of the interaction.
(For a review, see Ref. \cite{univers}.)
The presence of a harmonic trap introduces another parameter,
the trap frequency $\omega$ or, equivalently, the 
trap length $b=1/\sqrt{\mu \omega}$, where $\mu$ is
the reduced mass of two particles. As long as $b\gg r_0$,
the trapped system should still exhibit universal behavior,
which for $b \simle |a_2|$ could be significantly different 
from that of the untrapped system.
In the unitary limit $|a_2|\to \infty$, the untrapped two-body system
has a bound state at zero energy, and a collection 
of two-state fermions is characterized purely by the parameter that
sets the size of the system. 

While large-$|a_2|$ systems have been popular in 
atomic physics mainly in the last decade, they have been investigated 
in nuclear physics since its beginning. 
The two-nucleon ($NN$) 
system has two $S$-wave channels where $|a_2|\gg r_0$: 
the scattering lengths are about $5$ fm and $-20$ fm
in the $^3S_{1}$  and $^{1}S_{0}$ channels, respectively,
compared with a range of about $2$ fm. 
Here we 
combine two methods previously developed to deal with
untrapped particles ---effective field theory (EFT) and
no-core shell model (NCSM)--- in order to provide
solutions of few-fermion systems in a harmonic trap.

Since we are interested in the long-range dynamics, we can approximate
the complicated short-range physics as a series of contact interactions
that are delta functions with an increasing number of derivatives.
This can be formulated as a non-relativistic EFT
where observables are expressed in an expansion in powers of
$Qr_0$ (assuming, for simplicity, that the size of all scattering parameters
except for the scattering length is set by $r_0$).
The EFT in the untrapped two-body sector \cite{aleph} reproduces 
the effective-range expansion and is equivalent to a pseudopotential,
but can be extended to more-body systems \cite{3bodyEFT, 4bodyEFT}.
It can be shown that in leading order the Hamiltonian 
for two-component fermions
consists of 
a single, non-derivative, two-body contact interaction between the different
components.
For systems of identical bosons or more-component fermions,
a non-derivative, three-body contact interaction is also
present at this order.
The EFT for short-range forces and its generalization
for the exchange of light quanta are reviewed in
Ref. \cite{ARNPSreview}.

The NCSM is a powerful many-body technique for solving the 
Schr\"odinger equation for $A$ strongly interacting particles,
where the many-body basis states are constructed
using harmonic-oscillator (HO) wave functions.
In nuclear physics, NCSM is used to describe properties of light
nuclei without adjustable parameters 
\cite{Navratil:2000ww,Navratil:1999pwNavratil:2000gs}.
Starting from 
interactions that describe the
$NN$ scattering phase shifts and selected few-nucleon properties
with high accuracy, 
a unitary transformation is used 
to construct effective
interactions in truncated spaces of a finite number of energy excitations
on top of a minimum configuration. For the two-body problem, the
maximum number of excitations is equivalent 
to the number $N_{max}$  of shells included.
After a truncation where higher-body terms in the effective interaction
are neglected ---the so-called cluster
approximation---
numerical diagonalization allows 
a good description of nuclear energy spectra and other properties.
Unfortunately, there is no \textit{a priori} justification for
the cluster approximation in the context of phenomenological
interactions.

We have recently proposed \cite{NCSM_EFT} a combination of NCSM and EFT 
in which the cluster approximation is seen as a consequence of the EFT
power counting. 
Instead of performing a unitary transformation on phenomenological 
interactions,
we simply solve the EFT within the truncated space.
We determine its parameters from some binding energies
and then predict other bound-state properties by
extrapolation to the limits
$\omega \to 0$ and $N_{max}\to \infty$.
We have successfully applied the method in leading order to systems with 
$A\le 6$.
However, application beyond leading order
becomes cumbersome because of the increased number of parameters that
would have to be adjusted to properties of light nuclei. This motivates
us to devise a more flexible approach, which will allow us to determine
two-body parameters from two-body data. 

In this paper, we present the new method and apply it to the case
of two-component 
fermions in a harmonic trap.
The two \cite{Busch}
and three \cite{3bosluu:103202,unitgas_prl} -body systems 
with a pseudopotential in a trap
have already been studied. 
We provide an alternative, explicit solution for these systems in a NCSM basis 
where $\omega$ is kept fixed but $N_{max}$ is made large.
This allows us to test 
the accuracy of our approach in a non-trivial system.
In addition, we present a first solution of the four-body system.
The paper is organized as follows. We introduce  
our approach to the renormalization of the many-body problem
in Sec. \ref{sec:many_body},
and present our results in  Sec. \ref{sec:Results}.
Conclusions and perspectives for
future applications are discussed in Sec. \ref{sec:Concl}.

\section{The many-body problem and renormalization of the interaction
\label{sec:many_body}}

Here we consider a system of $A$ identical two-component 
fermions of mass $m$, where the 
two-body scattering length $a_2$ is large compared to 
the range $r_0$ of the interaction, $|a_2|\gg r_0$.
This system is assumed to be trapped in a harmonic potential
of frequency $\omega$,
whose length parameter 
is also sufficiently large, $b\gg r_0$.
We allow various values of the ratio $b/a_2$ ranging from the
unitary limit $b/|a_2|\to 0$ to the untrapped case $b/|a_2|\to \infty$.
(Note that we use units such that $\hbar=c=1$.)

Under these conditions, details of the inter-particle potential 
are irrelevant, and in leading order can be replaced by
a non-derivative two-body contact interaction between different components.
Without loss of generality we can refer to the two components 
as spin-1/2 states, in which case the contact interaction
acts only in the $^1S_0$ channel.
We denote its interaction strength by $C_0$. 
We start with the Hamiltonian for trapped particles,
\begin{equation}
H_A=\sum_{i=1}^{A}\left(\frac{p_{i}^{2}}{2m}
  +\frac{1}{2}m\omega^{2}r_{i}^{2}\right)
  +C_{0}\sum_{[i<j]}\delta^{(3)}(\vec{r}_{i}-\vec{r}_{j}),
\label{eq:Ham}
\end{equation}
where $\vec{r}_{i}$ and $\vec{p}_{i}$ represent the position and
momentum, respectively, of particle $i$, 
and $[i<j]$ denotes a pair of particles with opposite spin.
This Hamiltonian can be written as a sum of relative
and center-of-mass (CM) motion, so that the energy of the system is
the sum of internal ($E$) and CM ($E_{CM}$) terms:
\begin{equation}
E_{tot}=E_{CM}+E.
\label{eq:energycm}
\end{equation}
The CM motion is that of a simple HO, so in the following we focus
on the relative motion.

The internal Hamiltonian
is not well defined as it stands,
since the delta function is singular. Renormalization is necessary:
after a truncation of the Hilbert space with a cutoff in energy or momentum,
$C_0$ is taken to depend on the cutoff in such a way that 
finite, cutoff-independent results are obtained for observables.
The many-body problem cannot in general be solved analytically,
but 
renormalization can be incorporated easily in numerical
calculations, which are formulated from the outset within a finite space. 

These issues can be made explicit 
in the two-body system, where the relative motion of two
particles of opposite spin is described by
\begin{equation}
H_2=\frac{\omega}{2}\left[b^2 p^2+\frac{r^{2}}{b^2}
+2 \mu C_{0} b^2 \delta^{(3)}(\vec{r})\right]
\label{eq:rel2b}
\end{equation}
in terms of the relative coordinate $\vec{r}$ and the reduced mass $\mu=m/2$.

In the conventional NCSM approach, one chooses the model space by truncating
in the number of shells included in the calculation. More precisely,
one considers only HO states with the principal quantum number $N=2n+l$
smaller than a given value $N_{max}$. The effective Hamiltonian is
then constructed via a unitary transformation $U$ 
$(H_2^{eff}=U H_2 U^{\dagger})$,
designed so that the lowest $D$ eigenvalues of $H_2$ are exactly reproduced
by $H_2^{eff}$ ($D$ is the dimension of the model space). Of course,
the transformed Hamiltonian no longer has the form of the
initial interaction, and, in particular, additional non-local terms
are induced even if one starts with a contact interaction.

The alternative approach proposed in Ref. \cite{NCSM_EFT} is based
on principles of EFT, and it constructs the interaction between particles
making use of power counting. Thus, in each model space one preserves
the form of the interaction, as dictated by the power counting, and
one determines the strength of each interaction strength as
a function of $N_{max}$ and $\omega$
so that some observables are exactly reproduced. 
For example, one can fix  $C_0=C_0(N_{max}, \omega)$ such
that the two-body ground-state energy is fixed.

In this paper we propose an intermediate approach. Thus, as in 
Ref. \cite{NCSM_EFT}
we consider only the terms dictated by power counting, but instead
of adjusting their strength to reproduce observables in the few-body
system, we reproduce the $\omega$-dependent energies of two-body states,
as in the conventional NCSM approach. 
The number of states whose energies are to be reproduced
is fixed by the number of coupling constants to be adjusted, and not
by the dimension of the model space, as in the conventional approach
involving a unitary transformation. We note that, according to the
general principles of EFT, even if initially we start with only a
contact interaction, each truncation is going to induce additional
correction terms. Effectively, the truncation of the space induces
effective range, shape and other parameters, which
can be adjusted to the appropriate values as more derivatives
of delta functions are included.

To renormalize the interaction, we consider the relative Hamiltonian
(\ref{eq:rel2b}) and solve the corresponding Schr\"odinger equation,
\begin{equation}
\left[b^2 p^{2} +\frac{r^{2}}{b^2}
+2 \mu C_{0} b^2 \delta^{(3)}(\vec{r})\right]\psi(\vec{r})
=2 \frac{E}{\omega}\psi(\vec{r}),
\label{eq:contact2b}
\end{equation}
in a finite model space defined by $0\leq2n\leq N_{max}$. 
Since the delta function acts only on $S$ waves,
we consider here only those waves. 
The solution $\psi(\vec{r})$ can
be expanded in the complete set of HO wave functions 
\begin{equation}
\psi(\vec{r})=\sum_{n=0}^{N_{max}/2}A_{n}\phi_{n}(r),
\label{eq:wf_dec}
\end{equation}
with $\phi_{n}(r)$ the $S$-state solutions
of the unperturbed HO, 
\begin{equation}
\phi_{n}(r)=\frac{1}{\sqrt{4\pi}}
\left(\frac{2\: n!}{b^{3}\Gamma(n+3/2)}\right)^{1/2}
e^{-r^{2}/2b^{2}}L_{n}^{(1/2)}\left(\frac{r^{2}}{b^{2}}\right)
\label{eq:radwfs}
\end{equation}
in terms of generalized Laguerre polynomials.
One can easily verify that the eigenvalues of Eq. (\ref{eq:contact2b})
are given by the consistency condition
\begin{equation}
\frac{1}{C_{0}(N_{max},\omega)}=
-\sum_{n=0}^{N_{max}/2}\frac{|\phi_{n}(0)|^{2}}{(2n+3/2)\omega-E}.
\label{C0run}
\end{equation}

At this point, the coupling constant $C_{0}(N_{max},\omega)$
is undetermined. We can fix the value of $C_{0}(N_{max},\omega)$
at each $N_{max}$ by requiring that we reproduce a given
measured level $\tilde{E}(\omega)$ of
two trapped fermions \cite{stoferle:030401}.
However, the experimental values
are very close to the the theoretical prediction
of a pseudopotential, 
given \cite{Busch} by 
the transcendental equation
\begin{equation}
\frac{\Gamma(3/4-E/2\omega)}{\Gamma(1/4-E/2\omega)}
=\frac{b}{2a_2},
\label{eq:eigv_lo}
\end{equation}
where 
$a_2$ is the two-body
scattering length in the absence of the trap. 
(For discussions about the limits of applicability of
Eq. (\ref{eq:eigv_lo}), see Ref. \cite{models}).
The solutions of Eq. (\ref{eq:eigv_lo}) come in levels, and 
we can use any of the low-energy states to fix the value of 
$C_{0}(N_{max},\omega)$.
Here we use the most natural choice, the ground state.

Once we adjust the strength of the two-body term for each $N_{max}$
value (and, in general, $\omega$), we can perform few- and many-body
calculations. The many-body model spaces for $A>2$ are chosen so
that the two-body truncation is included consistently in relative
coordinates \cite{Navratil:2000ww}. 
Note that if one wants negative-parity states, 
one has to truncate the many-body space to odd $N_{max}$.
In this case, we use a two-body interaction that is adjusted to 
the largest even number below $N_{max}$, that is, $N_{max}-1$.

Results for the energies of few-fermion systems are in general complicated
functions of $N_{max}$ and $\omega$. 
In the unitary regime the dependence on $\omega$ gets simpler.
When $b/|a_2|\to 0$, the two-body spectrum of Eq. (\ref{eq:eigv_lo})
is given by $E_{n}/\omega=1/2 +2n$, where $n=0, 1, \ldots$.
Taking into account the $b$ dependence of Eq. (\ref{eq:radwfs}), 
we see from Eq. (\ref{C0run}) that 
the $\omega$ dependence factorizes:
$C_{0}(N_{max},\omega)= (2\pi b/ \mu) \, \gamma_{0}(N_{max})$,
where $\gamma_{0}(N_{max})$ is dimensionless and $\omega$ independent.
Indeed, in this case both $\mu$ and $b$ disappear from Eq. (\ref{eq:rel2b})
upon a rescaling $r \to b \rho$ (with $\rho$ a dimensionless variable),
and the energy can only be proportional to $\omega$.
More generally, rescaling all coordinates in Eq. (\ref{eq:Ham})
will ensure that 
$\omega$ exactly factors out in all the many-body eigenenergies. 
This was to be expected, given that for $|a_2|\to \infty$
the only energy parameter is $\omega$, which also sets the size of the system.
It has been argued \cite{genN} that the proportionality constant
between $E$ and $\omega$ is related to the short-distance scaling exponent
$\gamma$ of the wave function.
The three-fermion spectrum has been solved semi-analytically
in Ref. \cite{unitgas_prl}, while
the ground-state energies for $A=2\to 22$ have recently been calculated 
in the case of a short-range potential \cite{gfmc_unit}.

In the opposite limit $b/|a_2|\to \infty$, the trap is removed.
Eq. (\ref{eq:eigv_lo}) gives expected results.
There is a single two-body bound state of energy $E_0=-1/2\mu a_2^2$,
if the two-body interaction is sufficiently strong, $b/a_2\to \infty$.
All other states have energies $E_n/\omega= -1/2 + 2n$,
where $n=1,2,\ldots$, and correspond to scattering states.
When $b/a_2\to -\infty$, the interaction is weak
and the spectra of all few-body systems are expected to approach
those of non-interacting fermions in a HO. 
This can be seen from Eq. (\ref{C0run}):
as $b/a_2\to -\infty$ the two-body ground-state energy approaches
$3\omega/2$, leading to a divergence from the $n=0$ contribution and
to $C_0(N_{max}, \omega)\to 0$. 
In this limit all few-body energies should be set by occupation
of HO levels.
For example, all other
two-body levels are given by the poles in the right-hand side of 
Eq. (\ref{C0run}).
Whether few-body bound states exist when the interaction is sufficiently
attractive will be discussed below.

In any case, at the end of the calculation we extrapolate
to the limit $N_{max}\to \infty$. As we show in the next section,
this extrapolation is relatively smooth.
Our approach can be straightforwardly extended to higher orders.
At the two-body level, we add to Eq. (\ref{eq:Ham}) interactions proportional
to derivatives of the delta function, starting with an $S$-wave 
range correction of relative ${\cal O}(Qr_0)$ \cite{aleph}. 
A $P$-wave interaction (the most important
between like components) appears at ${\cal O}((Qr_0)^3)$.
For spin-1/2 fermions contact three- and more-body forces act in
$P$ and higher waves, so these forces are of relative 
${\cal O}((Qr_0)^5)$ or higher \cite{3bodyEFT}.
Many-body properties are thus to a high degree determined 
by two-body forces alone.
The two-body parameters can be determined from an extension of 
Eq. (\ref{eq:eigv_lo}) to non-negligible-range interactions \cite{models}.

\section{Results\label{sec:Results}}

In this section, we apply the method introduced in the previous section
to the description of systems of three and four identical fermions 
in a harmonic trap. 
We consider the unitary regime $(|a_{2}|\to\infty)$ as well
as a general non-vanishing $b/a_2$ value, for both positive and negative
scattering lengths.

In the three-body calculation, we have employed two numerical methods.
In the first stage, we have used the three-body
code in relative coordinates employed in Ref. \cite{NCSM_EFT}. Thus,
to handle three identical fermions of spin $1/2,$ we allow interaction
only in the $^{1}S_{0}$ channel, calculating the isospin $T=3/2$
solution. In a second stage, we have developed a new
program that can handle fermions of arbitrary spin, dropping the isospin
from the possible quantum numbers.
Both codes produce the same results for fermions of spin 1/2, but
the latter can be extended to larger $N_{max}$. We should point out that
in order to correctly antisymmetrize the three-body system, we include
all possible $l$ states in each model space. Therefore, although
we show below only results for the levels that are affected by the contact
interaction,
we also obtain energy
levels that are left unchanged. (For example,
states with $j^{\pi}=\frac{1}{2}^{+}$ can be obtained from addition of
two $l=1$ relative angular momenta and total spin $1/2$ or $3/2$;
however, in such channels the contact interaction is not present since it
appears only for $l=0$.)

\begin{figure}
\includegraphics[%
  clip,
  scale=0.70]{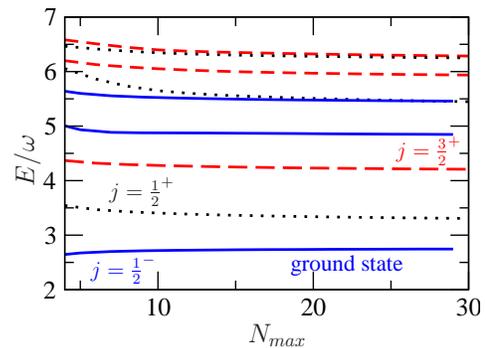}
\caption{(Color online) 
Dependence on the ultraviolet cutoff $N_{max}$ of
selected low-lying energies (in units of the HO frequency)
of three harmonically trapped fermions of spin $1/2$ 
in the unitary regime ($b/a_2=0$). We present
the lowest three states of $j^{\pi}=\frac{1}{2}^{-}$ (continuous
curve), $j^{\pi}=\frac{1}{2}^{+}$ (dotted curve), 
and $j^{\pi}=\frac{3}{2}^{+}$(dashed curve).\label{fig:lowsp_3b}}
\end{figure}

\begin{figure}
\includegraphics[%
  clip,
  scale=0.65]{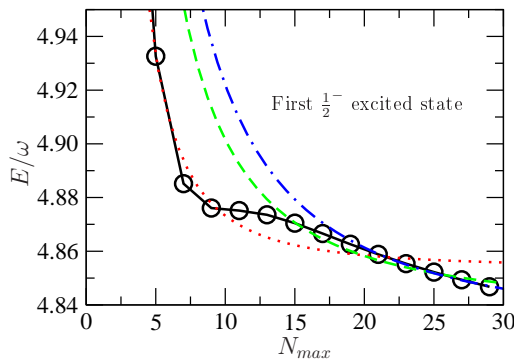}
\caption{(Color online) Running of the first $\frac{1}{2}^{-}$ excited state
(circles), and three fits with the functional form in Eq. (\ref{eq:fitform})
taking into account all points (dotted line), only points with
$N_{max}\geq15$ (dashed line) and only points with $N_{max}\geq 21$
(dot-dashed line). \label{fig:fitex}}
\end{figure}

Figure \ref{fig:lowsp_3b} presents the main results of our investigation
of the unitary regime. 
We show the dependence of the lowest-energy levels (labeled
by their spin and parity, $j^{\pi}$) in units of $\omega$
as function of the ultraviolet cutoff $N_{max}$.
We have verified numerically 
that, indeed, $\omega$
factorizes, as we have argued in the previous section.
The running of the low-lying energies is fairly smooth when one increases
the ultraviolet cutoff, and allows an extrapolation to
the limit $N_{max}\to\infty.$ However, we find that the running depends
on the state considered. In order to qualify this statement, we assume
a running of the form 
\begin{equation}
E=E_{\infty}+\frac{E_c}{(N_{max}+3/2)^{\alpha}},
\label{eq:fitform}
\end{equation}
with $E_{\infty}$, $E_c$, and $\alpha$ fitting parameters.
In a previous publication \cite{NCSM_EFT}, 
we assumed a running for the energy
of the many-body system of the form 
$E_{0}+E_1/\Lambda$, where $\Lambda=\sqrt{2\mu(N_{max}+3/2)\omega}$
was defined as the ultraviolet cutoff. (Because here the solutions
depend trivially on $\omega$, we do not have to include it in
the definition of the cutoff.) This was motivated by the running in
the continuum two-body system. 
Here we carry out a more detailed investigation
of the running of the three-body solution with the ultraviolet cutoff. 
Allowing $\alpha$ to vary in the fit, we find values between $0.5$ and $2.0$.
We find that, in general, the values of $E_{\infty}$ are not very
sensitive to the value of $\alpha$, but can
depend on the features of the running and the values of $N_{max}$
included in the fit. 
For example, while in Fig. \ref{fig:lowsp_3b}
the first $j^{\pi}=\frac{1}{2}^{-}$ excited state appears to be smooth,
a closer examination reveals some structure, as shown in Fig. \ref{fig:fitex}.
In this figure
we exhibit also three possibilities for fitting, 
where we include all points, 
only points with $N_{max}\geq 15$, or only points with $N_{max}\geq 21$. 
Since Eq. (\ref{eq:fitform})
is valid for large $N_{max}$, our best guess for $\alpha$ is obtained
in the case when we consider only $N_{max}\geq 21$.
In this case we find $\alpha \simeq 1.474$, and
the extrapolated value for the energy of the
state is $E_\infty/\omega \simeq 4.83$. 

In Tables \ref{cap:j12m}--\ref{cap:j32p}
we present our estimates for $E_\infty$ for the states shown
in Fig. \ref{fig:lowsp_3b}.
In our approach, there are two sources of errors: 
on one hand, we neglect higher-order terms in the expansion of the 
interaction (terms involving derivatives of
contact interactions) and, on the other hand, we perform a fit of the form of
Eq. (\ref{eq:fitform}) 
to obtain the large-$N_{max}$ limit.
The variation with $N_{max}$ allows us to estimate errors associated
with missing terms; thus, for
low-lying states, even fairly small values of $N_{max}$ produce
results within about $10\%$ of the extrapolated values. On the other hand,
we find that, in
general, the extrapolation errors coming from fits to large 
values of $N_{max}$
are very small.
For the levels shown in Tables \ref{cap:j12m}--\ref{cap:j32p}
these extrapolation errors are beyond the last digit shown, 
except for 
second level in Table \ref{cap:j12m}, 
where the error is 2 in the last digit.

In the limit $N_{max}\to\infty$,
we should recover
the semi-analytical solutions obtained in Ref. \cite{unitgas_prl} 
under the assumption of a two-body pseudopotential: 
\begin{equation}
\frac{E}{\omega}= 1+s+2q,
\label{eq:energ_3tf12}
\end{equation}
where $q=0, 1, \ldots$, and $s\geq 1.77$ is the (real) solution of 
a transcendental equation. 
The agreement between our numerical method and the semi-analytical values,
also shown in Tables \ref{cap:j12m}--\ref{cap:j32p},
is remarkable and provides a confirmation of the reliability of our method.
As a side result, we note that the short-range scaling exponent \cite{genN}
we obtain for three particles is $\gamma\simeq-0.24$.  
(While the
virial theorem \cite{genN} is not satisfied necessarily in each model space, 
we expect it
to be satisfied in the large-$N_{max}$ limit.)

\begin{table}
\caption{Comparison between the results of the present approach 
($E_\infty/\omega$)
and of the semi-analytical formula from Ref. \cite{unitgas_prl}
(Eq. (\ref{eq:energ_3tf12})), 
for \label{cap:j12m} $j^{\pi}=\frac{1}{2}^{-}$.}
\begin{tabular}{cccccc}
\hline \hline
$n$&
$l$&
$q$&
$s$&
Eq. (\ref{eq:energ_3tf12})&
$E_\infty/\omega$\tabularnewline
\hline 
0&
1&
0&
1.77&
2.77&
2.76\tabularnewline
0&
1&
1&
1.77&
4.77&
4.71\tabularnewline
1&
1&
0&
4.36&
5.36&
5.39\tabularnewline
\hline \hline
\end{tabular}
\end{table}

\begin{table}
\caption{Same as in Table \ref{cap:j12m}, but for $j^{\pi}=\frac{1}{2}^{+}.
$\label{cap:j12p}}
\begin{tabular}{cccccc}
\hline \hline
$n$&
$l$&
$q$&
$s$&
Eq. (\ref{eq:energ_3tf12})&
$E_\infty/\omega$\tabularnewline
\hline
0&
0&
0&
2.17&
3.17&
3.17\tabularnewline
0&
0&
1&
2.17&
5.17&
5.13\tabularnewline
1&
0&
0&
5.13&
6.13&
6.15\tabularnewline
\hline \hline
\end{tabular}
\end{table}

\begin{table}
\caption{Same as in Table \ref{cap:j12m}, but for $j^{\pi}=\frac{3}{2}^{+}.
$\label{cap:j32p}}
\begin{tabular}{cccccc}
\hline \hline
$n$&
$l$&
$q$&
$s$&
Eq. (\ref{eq:energ_3tf12})&
$E_\infty/\omega$\tabularnewline
\hline
0&
2&
0&
3.10&
4.10&
4.11\tabularnewline
1&
2&
0&
4.79&
5.79&
5.81\tabularnewline
0&
2&
1&
3.10&
6.10&
6.07\tabularnewline
\hline \hline
\end{tabular}
\end{table}

The successful test in the unitary regime encourages us to extend the
application of our method for arbitrary values of the $b/a_2$ ratio.
Experimentally, one can vary the scattering length of trapped particles 
by means
of a magnetic field, and, in principle, obtain a large range of ratios 
$b/a_2$.
Thus, for illustration, we have considered 
a range of $b/a_2$ values,
and fixed the coupling constant in each model space so that the ground-state
energy given by Eq. (\ref{eq:eigv_lo}) is always exactly reproduced.

\begin{figure}[t]
\includegraphics[%
  clip,
  scale=0.7]{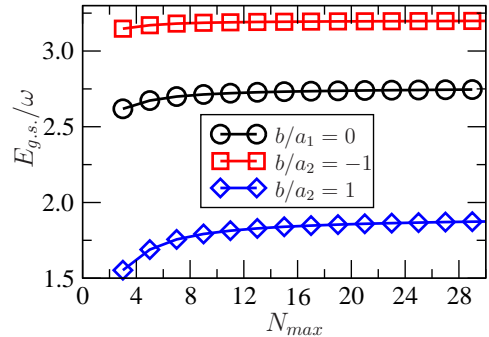}
\caption{(Color online) Ground-state energies (in units of the HO frequency)
of the three-fermion system in
a harmonic trap as function of the cutoff $N_{max}$
for $b/a_2=0$ (circles), $b/a_2=-1$
(squares) and $b/a_2=1$ (diamonds). }\label{fig:gs3f}
\end{figure}

In Fig. \ref{fig:gs3f} we show the running of
the lowest $j^\pi=\frac{1}{2}^-$  (ground state)
energy of the trapped
three-fermion system for
$b/a_2=\pm 1$ in comparison with the unitary limit. 
We obtain different shifts of the energy level with respect
to the unitary solution depending on the sign of $a_2$. 
If $a_2>0,$ the level is
pushed downward in energy with respect to the unitary value, 
while if $a_2<0$
the level is pushed upward. This behavior
is a reflection of the same type of behavior in the two-body system:
while in the unitary limit the two-body ground state is 
$E_{0}/\omega=1/2$,
the corresponding energy for $b/a_2=1$ 
decreases to $E_{0}/\omega\simeq -0.34$,
and for  $b/a_2=-1$ increases to $E_{0}/\omega\simeq 0.9$.

\begin{figure}[t]
\includegraphics[%
  clip,
  scale=0.7]{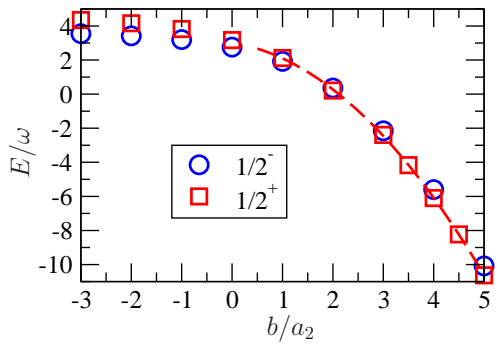}
\caption{(Color online) Lowest (extrapolated) energy levels for 
$j^\pi=\frac{1}{2}^-$ (circles) and 
$j^\pi=\frac{1}{2}^+$ (squares) states of the three-fermion system in
a harmonic trap as function of the the ratio $b/a_2$. Around $b/a_2\sim 1.5$,
we observe an inversion of the parity of the system's ground state.  
The
dashed line is a fit of the positive-parity points by a quadratic form.
\label{fig:gs3f_ba}}
\end{figure}

In Fig. \ref{fig:gs3f_ba}
we consider both $j^\pi=\frac{1}{2}^-$ and $j^\pi=\frac{1}{2}^+$ states, and
plot the
lowest-energy level in each case as a function of $b/a_2$. 
We show only the extrapolated values $E_\infty$ obtained by means of Eq.  
(\ref{eq:fitform}), taking into account only the points with 
$N_{max}\geq 21$ to reduce fitting errors. 
For $b/a_2 \simle 1.5$, the ground state is the same as in the unitary 
limit.
However, we observe a tendency for an inversion of the parity of 
the ground state for $b/a_2\simge 1.5$, although 
the positive- and negative-parity levels are nearly degenerate.
Note that in the region with $a_2>0$ 
we might have larger errors from the fitting procedure, 
because the interaction strength increases
with increasing values of the ratio $b/a_2$.

As discussed in the previous section, for weak two-body attraction the
spectrum should approach that of three non-interacting fermions in a HO trap
---which is itself wide. 
In the limit $b/a_2\to-\infty$, the lowest state is
a configuration with two particles in the first $S$ state, 
and the third in the first $P$ level.
Figure \ref{fig:gs3f_ba} indeed suggests the asymptotic value
$4\omega$ ($5 \omega$) for the energy of
the negative (positive) -parity state.

On the other hand, for strong two-body attraction the relative effect of
the contact interaction increases. 
In order to understand the possible parity inversion,
we fit the $a_2>0$ results for $E/\omega$ in Fig. \ref{fig:gs3f_ba}
to a quadratic function of $b/a_2$ (dashed line).
In the untrapped $b/a_2\to\infty$ limit 
we find that the lowest
positive- and negative-parity levels are very close, and their energy is
\begin{equation}
E\simeq -\frac{1}{2\mu a^2_2}.
\end{equation}
Thus, within errors, the system of three fermions 
is at the threshold for the scattering of one particle
on the bound state of the other two (dimer):
the system is not bound, 
or very weakly bound, and can be
viewed as composed of a dimer 
and an additional particle.
The lack of a deep three-body bound state 
follows in fact from a very naive argument.
The existence of a shallow two-body bound state is a consequence
of a balance between the renormalized delta-function attraction
and the kinetic repulsion. The addition of a third spin-1/2 fermion
roughly doubles both, so we do not expect a collapse
of the three-body system ---as one does when the number of pair interactions 
grows
faster than the number of particles, 
e.g. for identical bosons and multi-component fermions. 

The parity inversion of the ground state is therefore plausible.
In the $b/a_2\to \infty$ limit,
the positive-parity state 
likely represents
the untrapped $S$-wave particle-dimer scattering state that one expects to 
dominate sufficiently
close to threshold.
This is consistent with results \cite{petrov3}
for near-threshold particle-dimer scattering.

We have also solved the four-fermion problem in the trap.
For now, we are limited
to smaller model spaces than for three fermions because we use the
\textsc{redstick} Slater determinant code \cite{redstick} that preserves
only the third component of the angular-momentum projection,
mixing states with good angular momentum. The dimension of
the many-body basis thus increases significantly. Because we truncate the
many-body configurations allowing only a maximum number of excitations
on top of the minimum solution, we can eliminate spurious CM contributions
when we compute the spectrum of the intrinsic motion. (The calculations
in relative coordinates and in a properly truncated Slater determinant
basis are, in fact, equivalent.) Moreover, because the Hamiltonian
is rotationally invariant, we obtain eigenstates with good angular
momentum (for four particles, the lowest positive-parity state has $j=0$), 
even though
most of the individual Slater determinants that compose the many-body
basis do not have good $j$. 

\begin{figure}[t]
\includegraphics[%
  clip,
  scale=0.7]{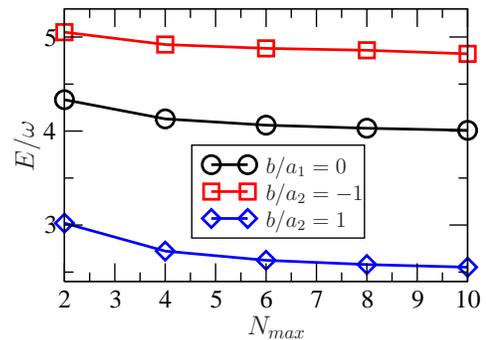}
\caption{(Color online) Same as Fig. \ref{fig:gs3f} 
but for the $j^\pi=0^+$ states
of the four-fermion system.
\label{fig:gs4f}}
\end{figure}

Figure \ref{fig:gs4f} exhibits the running of the 
lowest positive-parity levels 
of four harmonically trapped spin-1/2 fermions, with the same parameters as
in Fig. \ref{fig:gs3f}, $b/a_2=0, \pm 1$.
Because of the present limitation to just a few values of $N_{max}$ for the
four-body system,  our fitting procedure is
likely to be subject to larger extrapolation errors, although
Fig. \ref{fig:gs4f} shows small variation with the ultraviolet cutoff.
Thus, in the unitary limit, 
if in Eq. (\ref{eq:fitform}) we use $\alpha=1/2$, we estimate 
$E_{\infty}/\omega\sim 3.58$, while if we use $\alpha=1$, we obtain
$E_{\infty}/\omega\sim 3.85$. (The small number of points does not allow
us to obtain $\alpha$ reliably from the fit.) 
Both values are in line with a previous
calculation of the four-body system in the unitary
limit \cite{gfmc_unit}. 
{}From these results we can estimate the short-range
scaling coefficient $\gamma$ \cite{genN} for four particles
between  $-0.65$ and $-0.92$.

At non-vanishing $b/a_2$ we observe the same type of displacement 
in the lowest 
positive-parity level as for two- and three-fermion systems. 
However, our technical limitations in the four-body system do not allow us yet
to make a more detailed investigation of a large range of ratios
$b/a_2$. 
As expected, the system goes to the non-interacting case in the limit
$b/a_2\to -\infty$, that is, the ground-state energy is $13\omega/2$.
As for three particles, we can extrapolate 
to the other limit $b/a_2\to \infty$ with a quadratic form in $b/a_2>0$. 
We find the binding energy in this limit to be
about 40\% below the threshold of scattering of two dimers.
However, because our results are subject to large errors, 
we leave a more thorough analysis of four-body
spectra in different regimes for future work
using relative coordinates, 
when it should be
possible to achieve larger ultraviolet
cutoffs.

\section{Conclusions and outlook \label{sec:Concl}}

We have presented a first application of the NCSM to the description
of three- and four-fermion systems in a harmonic trap. 
Based on the general
principles of EFT, we have proposed a method to renormalize the two-body
interaction to be used for few- and many-body calculations in finite
model spaces. Tests of our results against semi-analytical calculations
in the unitary regime suggest good accuracy in estimation of
the energy levels of the three-fermion system. Moreover, we have extended
the application of our method to arbitrary $b/a_2$ ratios.

There are, of course, ways to improve our results. In principle, even
if a pseudopotential is a good approximation, in finite model spaces
we can add a range correction to the contact term. This should
improve the running of observables with the ultraviolet cutoff. Although
in many cases the running is quite small, we could improve
the calculations for $a_2>0$, where the interaction is stronger
and requires larger model spaces to converge. 
(In the four-nucleon
system we can follow the evolution of the contribution of different
configurations. We find that the components with the lowest-energy
configuration dominate for $a_2<0$, while other configurations
reshuffle small contributions in each model space. However, this situation
changes, with a larger contribution of the higher shells, for a stronger
interaction such as for $a_2>0$. This is not surprising
since a stronger
interaction becomes more important across shells.) Adding higher-order 
terms should decrease the magnitude of $E_c$ in Eq. (\ref{eq:fitform}),
thereby allowing for a more precise determination of observables in
few- and many-body systems.

In principle, we could extend our method even further. First, we can
perform similar calculations of other observables (e.g., r.m.s. radii), 
with more particles, as well as for fermions of different
spin. Especially for the latter it will be interesting to investigate
the ability of the method to predict Efimov states in a finite model
space. In the case of untrapped multi-state fermions,
the EFT power
counting requires a contact three-body force to prevent the collapse
of the three-body system \cite{3bodyEFT}.

Finally, the same type of approach can be considered in nuclear physics,
where addition of a CM harmonic term to the intrinsic Hamiltonian
produces effectively the same type of two-body relative Hamiltonian
as in Eq. (\ref{eq:rel2b}) \cite{Lipkin_CM}.

\section*{Acknowledgments}

We thank J. Carlson for useful discussions,
H. Hammer and T. Papenbrock for pointing out useful references, and
F. Werner and Y. Castin for providing their values for energies
shown in Tables \ref{cap:j12m}--\ref{cap:j32p}. 
This research was supported in part
by the U.S. Department of Energy under grant numbers
DE-AC53-06NA25396 (IS), DE-FG02-04ER41338 (UvK), and 
DE-FG02-87ER40371 (JPV),
by the U.S. National Science Foundation under grant PHY0555396 (BRB),
by the Nederlandse Organisatie voor Wetenschappelijk Onderzoek  (UvK),
and by Brazil's FAPESP under a Visiting Professor grant (UvK).
UvK would like to thank the hospitality of 
the Kernfysisch Versneller Instituut at Rijksuniversiteit Groningen,
the Instituto de F\'\i sica Te\'orica of the
Universidade Estadual Paulista,
the Instituto de F\'\i sica of the Universidade de S\~ao Paulo,
and the Institute for Nuclear Theory at the University of Washington,
where part of this work was carried out.

%\bibliographystyle{apsrev}
%\bibliography{myreferences}

\end{document}